\documentclass[twocolumn,showpacs,preprintnumbers,superscriptaddress,amsmath,amssymb,PRC]{revtex4}
\usepackage{graphicx}
\usepackage{dcolumn}
\usepackage{float}
\usepackage{mathptmx, courier, pifont}
\usepackage[scaled=0.92]{helvet}
\usepackage[T1]{fontenc}
\usepackage{textcomp}
\usepackage{color}

\begin{document}
\title{Isochronous mass measurements of $T_z=-1$ $fp$-shell nuclei from projectile fragmentation of $^{58}$Ni}
\author{Y.~H.~Zhang}
\affiliation{CAS Key Laboratory of High Precision Nuclear Spectroscopy and Center for Nuclear Matter Science, Institute of Modern Physics, Chinese Academy of Sciences, Lanzhou 730000, People's Republic of China}
\affiliation{Joint Research Center for Modern Physics and Clean Energy, South China Normal University, Institute of Modern Physics, Chinese Academy of Sciences, Lanzhou 730000, People's Republic of China}
\author{P.~Zhang}
\affiliation{CAS Key Laboratory of High Precision Nuclear Spectroscopy and Center for Nuclear Matter Science, Institute of Modern Physics, Chinese Academy of Sciences, Lanzhou 730000, People's Republic of China}
\affiliation{University of Chinese Academy of Sciences,
	Beijing, 100049, People's Republic of China}
\author{X.~H.~Zhou}\thanks{Corresponding author. Email address: zxh@impcas.ac.cn}
\affiliation{CAS Key Laboratory of High Precision Nuclear Spectroscopy and Center for Nuclear Matter Science, Institute of Modern Physics, Chinese Academy of Sciences, Lanzhou 730000, People's Republic of China}
\affiliation{Joint Research Center for Modern Physics and Clean Energy, South China Normal University, Institute of Modern Physics, Chinese Academy of Sciences, Lanzhou 730000, People's Republic of China}
\author{M.~Wang}\thanks{Corresponding author. Email address: Wangm@impcas.ac.cn}
\affiliation{CAS Key Laboratory of High Precision Nuclear Spectroscopy and Center for Nuclear Matter Science, Institute of Modern Physics, Chinese Academy of Sciences, Lanzhou 730000, People's Republic of China}
\affiliation{Joint Research Center for Modern Physics and Clean Energy, South China Normal University, Institute of Modern Physics, Chinese Academy of Sciences, Lanzhou 730000, People's Republic of China}
\author{Yu.~A.~Litvinov}\thanks{Corresponding author. Email address: y.litvinov@gsi.de}
\affiliation{CAS Key Laboratory of High Precision Nuclear Spectroscopy and Center for Nuclear Matter Science, Institute of Modern Physics, Chinese Academy of Sciences, Lanzhou 730000, People's Republic of China}
\affiliation{GSI Helmholtzzentrum f\"{u}r Schwerionenforschung,
	Planckstra{\ss}e 1, 64291 Darmstadt, Germany}
\author{H.~S.~Xu}
\affiliation{CAS Key Laboratory of High Precision Nuclear Spectroscopy and Center for Nuclear Matter Science, Institute of Modern Physics, Chinese Academy of Sciences, Lanzhou 730000, People's Republic of China}
\affiliation{Joint Research Center for Modern Physics and Clean Energy, South China Normal University, Institute of Modern Physics, Chinese Academy of Sciences, Lanzhou 730000, People's Republic of China}
\author{X.~Xu}
\affiliation{CAS Key Laboratory of High Precision Nuclear Spectroscopy and Center for Nuclear Matter Science, Institute of Modern Physics, Chinese Academy of Sciences, Lanzhou 730000, People's Republic of China}
\author{P.~Shuai}
\affiliation{CAS Key Laboratory of High Precision Nuclear Spectroscopy and Center for Nuclear Matter Science, Institute of Modern Physics, Chinese Academy of Sciences, Lanzhou 730000, People's Republic of China}
\author{Y.~H.~Lam}
\affiliation{CAS Key Laboratory of High Precision Nuclear Spectroscopy and Center for Nuclear Matter Science, Institute of Modern Physics, Chinese Academy of Sciences, Lanzhou 730000, People's Republic of China}
\author{R.~J.~Chen}
\affiliation{CAS Key Laboratory of High Precision Nuclear Spectroscopy and Center for Nuclear Matter Science, Institute of Modern Physics, Chinese Academy of Sciences, Lanzhou 730000, People's Republic of China}
\affiliation{GSI Helmholtzzentrum f\"{u}r Schwerionenforschung,
	Planckstra{\ss}e 1, 64291 Darmstadt, Germany}
\author{X.~L.~Yan}
\affiliation{CAS Key Laboratory of High Precision Nuclear Spectroscopy and Center for Nuclear Matter Science, Institute of Modern Physics, Chinese Academy of Sciences, Lanzhou 730000, People's Republic of China}
\author{T.~Bao}
\affiliation{CAS Key Laboratory of High Precision Nuclear Spectroscopy and Center for Nuclear Matter Science, Institute of Modern Physics, Chinese Academy of Sciences, Lanzhou 730000, People's Republic of China}
\author{X.~C.~Chen}
\affiliation{CAS Key Laboratory of High Precision Nuclear Spectroscopy and Center for Nuclear Matter Science, Institute of Modern Physics, Chinese Academy of Sciences, Lanzhou 730000, People's Republic of China}
\affiliation{Max-Planck-Institut f\"{u}r Kernphysik, Saupfercheckweg 1, 69117 Heidelberg, Germany}
\author{H.~Chen}
\affiliation{CAS Key Laboratory of High Precision Nuclear Spectroscopy and Center for Nuclear Matter Science, Institute of Modern Physics, Chinese Academy of Sciences, Lanzhou 730000, People's Republic of China}
\affiliation{University of Chinese Academy of Sciences,
	Beijing, 100049, People's Republic of China}
\author{C.~Y.~Fu}
\affiliation{CAS Key Laboratory of High Precision Nuclear Spectroscopy and Center for Nuclear Matter Science, Institute of Modern Physics, Chinese Academy of Sciences, Lanzhou 730000, People's Republic of China}
\affiliation{University of Chinese Academy of Sciences,
	Beijing, 100049, People's Republic of China}
\author{J.~J.~He}
\affiliation{CAS Key Laboratory of Optical Astronomy, National Astronomical Observatories, Chinese Academy of Sciences, Beijing, 100012, People's Republic of China}
\affiliation{CAS Key Laboratory of High Precision Nuclear Spectroscopy and Center for Nuclear Matter Science, Institute of Modern Physics, Chinese Academy of Sciences, Lanzhou 730000, People's Republic of China}
\author{S.~Kubono}
\affiliation{CAS Key Laboratory of High Precision Nuclear Spectroscopy and Center for Nuclear Matter Science, Institute of Modern Physics, Chinese Academy of Sciences, Lanzhou 730000, People's Republic of China}
\author{D.~W.~Liu}
\affiliation{CAS Key Laboratory of High Precision Nuclear Spectroscopy and Center for Nuclear Matter Science, Institute of Modern Physics, Chinese Academy of Sciences, Lanzhou 730000, People's Republic of China}
\affiliation{University of Chinese Academy of Sciences,
	Beijing, 100049, People's Republic of China}
\author{R.~S.~Mao}
\affiliation{CAS Key Laboratory of High Precision Nuclear Spectroscopy and Center for Nuclear Matter Science, Institute of Modern Physics, Chinese Academy of Sciences, Lanzhou 730000, People's Republic of China}
\author{X.~W.~Ma}
\affiliation{CAS Key Laboratory of High Precision Nuclear Spectroscopy and Center for Nuclear Matter Science, Institute of Modern Physics, Chinese Academy of Sciences, Lanzhou 730000, People's Republic of China}
\affiliation{Joint Research Center for Modern Physics and Clean Energy, South China Normal University, Institute of Modern Physics, Chinese Academy of Sciences, Lanzhou 730000, People's Republic of China}
\author{M.~Z.~Sun}
\affiliation{CAS Key Laboratory of High Precision Nuclear Spectroscopy and Center for Nuclear Matter Science, Institute of Modern Physics, Chinese Academy of Sciences, Lanzhou 730000, People's Republic of China}
\affiliation{University of Chinese Academy of Sciences,
	Beijing, 100049, People's Republic of China}
\author{X.~L.~Tu}
\affiliation{CAS Key Laboratory of High Precision Nuclear Spectroscopy and Center for Nuclear Matter Science, Institute of Modern Physics, Chinese Academy of Sciences, Lanzhou 730000, People's Republic of China}
\affiliation{Joint Research Center for Modern Physics and Clean Energy, South China Normal University, Institute of Modern Physics, Chinese Academy of Sciences, Lanzhou 730000, People's Republic of China}
\affiliation{Max-Planck-Institut f\"{u}r Kernphysik, Saupfercheckweg 1, 69117 Heidelberg, Germany}
\author{Y.~M.~Xing}
\affiliation{CAS Key Laboratory of High Precision Nuclear Spectroscopy and Center for Nuclear Matter Science, Institute of Modern Physics, Chinese Academy of Sciences, Lanzhou 730000, People's Republic of China}
\affiliation{University of Chinese Academy of Sciences,
	Beijing, 100049, People's Republic of China}
\affiliation{GSI Helmholtzzentrum f\"{u}r Schwerionenforschung,
	Planckstra{\ss}e 1, 64291 Darmstadt, Germany}
\author{Q.~Zeng}
\affiliation{Research Center for Hadron Physics, National Laboratory of Heavy Ion Accelerator Facility in Lanzhou and University of Science and Technology of China, Hefei 230026, People's Republic of China}
\affiliation{CAS Key Laboratory of High Precision Nuclear Spectroscopy and Center for Nuclear Matter Science, Institute of Modern Physics, Chinese Academy of Sciences, Lanzhou 730000, People's Republic of China}
\author{X.~Zhou}
\affiliation{CAS Key Laboratory of High Precision Nuclear Spectroscopy and Center for Nuclear Matter Science, Institute of Modern Physics, Chinese Academy of Sciences, Lanzhou 730000, People's Republic of China}
\affiliation{University of Chinese Academy of Sciences,
	Beijing, 100049, People's Republic of China}
\author{W.~L.~Zhan}
\affiliation{CAS Key Laboratory of High Precision Nuclear Spectroscopy and Center for Nuclear Matter Science, Institute of Modern Physics, Chinese Academy of Sciences, Lanzhou 730000, People's Republic of China}
\author{S.~Litvinov}
\affiliation{GSI Helmholtzzentrum f\"{u}r Schwerionenforschung,
	Planckstra{\ss}e 1, 64291 Darmstadt, Germany}
\author{K.~Blaum}
\affiliation{Max-Planck-Institut f\"{u}r Kernphysik, Saupfercheckweg 1, 69117 Heidelberg, Germany}
\author{G.~Audi}
\affiliation{CSNSM-IN2P3-CNRS, Universit\'{e} de Paris Sud, F-91405
	Orsay, France}
\author{T.~Uesaka}
\affiliation{RIKEN Nishina Center, RIKEN, Saitama 351-0198, Japan}
\author{Y. Yamaguchi}
\affiliation{RIKEN Nishina Center, RIKEN, Saitama 351-0198, Japan}
\author{T. Yamaguchi}
\affiliation{Department of Physics, Saitama University, Saitama 338-8570, Japan}
\author{A.~Ozawa}
\affiliation{Insititute of Physics, University of Tsukuba, Ibaraki 305-8571, Japan}
\author{B.~H.~Sun}
\affiliation{School of Physics and Nuclear Energy Engineering,
	Beihang University, Beijing 100191, People's Republic of China}
\author{Y.~Sun}
\affiliation{Department of Physics and Astronomy, Shanghai Jiao Tong University,
	Shanghai 200240, People's Republic of China} 
\author{F.~R.~Xu}
\affiliation{State Key Laboratory of Nuclear Physics and Technology, School of Physics, Peking University, Beijing 100871, People's Republic of China} 



\begin{abstract}
	
Atomic masses of seven $T_z=-1$, $fp$-shell nuclei from $^{44}$V to $^{56}$Cu and two low-lying isomers, $^{44m}$V ($J^\pi=6^+$) and $^{52m}$Co ($J^\pi=2^+$), have been measured with relative precisions of $1-4\times 10^{-7}$ 
with Isochronous Mass Spectrometry (IMS) at CSRe. The masses of $^{56}$Cu, $^{52g,52m}$Co, and $^{44m}$V were measured for the first time in this experiment. The Mass Excesses ($ME^{\prime}$s) of $^{44}$V, $^{48}$Mn, $^{50}$Fe, and $^{54}$Ni are determined with an order of magnitude improved precision compared to the literature values. $^{52g,52m}$Co and $^{56}$Cu are found to be $370$~keV and $400$~keV more bound, respectively, while $^{44g,44m}$V are $\sim 300$~keV less bound than the extrapolations in the Atomic-Mass Evaluation 2012 (AME$^{\prime}$12). The masses of the four $T_z=-1/2$ nuclei $^{45}$V, $^{47}$Cr, $^{49}$Mn, and $^{51}$Fe 
are re-determined to be in agreement, within the experimental errors, with the recent JYFLTRAP measurements or with the previous IMS measurements in CSRe. Details of the measurements and data analysis are described, and the impact of the new $ME$ values on different aspects in nuclear structure are investigated and discussed.
	
\end{abstract}

\pacs{21.10.Dr, 27.40.+z, 29.20.db}

\maketitle



\section{Introduction}

Nuclear mass measurements provide information on nuclear binding energies which reflect the summed results of all interactions among its constituent protons and neutrons. The systematic and accurate knowledge of nuclear masses have wide applications in many areas of subatomic physics ranging from nuclear structure and astrophysics to fundamental interactions and symmetries depending on the mass precision achieved~\cite{Blaum06,Lunney03}. For example, on the basis of nuclear masses, the well-known shell structure and pairing correlations were discovered in stable nuclei~\cite{Bohr98}, and the disappearance of the magic neutron number $N=20$~\cite{Thi75} as well as the new shell closure at $N=32$~\cite{Wien13} were revealed in exotic neutron-rich nuclides. In addition to the mapping of the nuclear mass surface~\cite{Rad00,Nov02,Sch13}, much attention has been paid to precision mass measurements of exotic nuclei in specific mass regions, such as in the vicinity of doubly magic nuclei far from stability and the waiting point nuclei in the rapid proton and rapid neutron capture processes (see Refs.\cite{Blaum06,Lunney03} for reviews). 
\par
The interest in precision mass measurements for the neutron-deficient $fp$-shell nuclei are due to several considerations: (1) These nuclides are located at the reaction path of the nucleosynthesis  $rp$-process~\cite{wal81} in X-ray bursts; the corresponding $(p,\gamma)$ reaction $Q$ values, deduced from masses of the nuclides involved, are key nuclear physics inputs in the model calculations~\cite{Sch06,Par09}. (2) Precise masses have been used for testing the Isobaric Multiplet Mass Equation (IMME) (see Refs.~\cite{Mac14,Lyh13} for reviews),
an important issue associated with isospin symmetry in particle and nuclear physics. (3) The precise $Q_{\rm EC}$ values, deduced from the mass differences of parent and daughter nuclei, are required in the $\beta$-decay studies in order to obtain the Fermi (F) and Gamow-Teller (GT) transition strengths. In particular, 
the high precision $Q_{\rm EC}$ values of super-allowed $0^+ \rightarrow 0^+$ Fermi decays provide one of the most important quantities required for testing the Conserved Vector Current (CVC) hypothesis~\cite{TH15b}. 
(4) The experimental masses are used to predict, with the help of local mass relationships such as the IMME and Garvey-Kelson (GK) mass formula~\cite{Gar66}, the mass values of more neutron-deficient nuclei, which in turn are essential for understanding
the astrophysical $rp$-process of nucleosynthesis~\cite{Sch06,Par09} and identifying the most probable candidates for two-proton radioactivity.     
\par
In the past few years, a series of mass-measurement experiments have been performed using Isochronous Mass Spectrometry (IMS) in the Cooler Storage Ring (CSRe), Lanzhou~\cite{Zhang16}. In this article, we report on the measured atomic masses of $^{58}$Ni projectile fragments focusing on the $T_z=-1$ short-lived $fp$-shell nuclei. A part of the obtained results has been discussed in our previous publications~\cite{Xu16,Zhang17}. Details of the experimental measurements and data analysis are described in Section II. The experimental results are presented in Section III. The impact of our results in nuclear structure studies
is discussed in Section IV. We give a summary and conclusion in Section V. 
\section{Experiment and data analysis}
\subsection{Measurement}
\begin{figure} [t]
	\begin{center}
		\includegraphics[angle=0,width=8.cm]{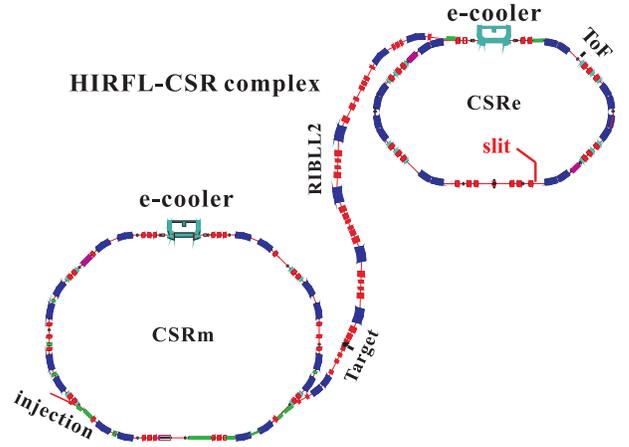}
		\caption{(Colour online)
			Layout of the high-energy part of the HIRFL-CSR complex at IMP including the synchrotron CSRm, the in-flight fragment separator
			RIBLL2, and the experimental storage ring CSRe. 
		} \label{Fig01}
	\end{center}
\end{figure}

The experiment was conducted at the HIRFL-CSR acceleration complex~\cite{Xia02,zwl2010}, which consists of a Separated Sector Cyclotron (SSC, $K=450$), a Sector-Focusing Cyclotron (SFC, $K=69$), a main Cooler-Storage Ring (CSRm) operating as a heavy-ion synchrotron, and an experimental storage ring CSRe. The two storage rings are coupled by an in-flight fragment separator RIBLL2. The high-energy part of the facility is schematically shown in Fig.~\ref{Fig01}. The CSRm has a circumference of 161 m and a maximum magnetic rigidity $B\rho=12.05$ Tm. The $^{12}$C$^{6+}$ and $^{238}$U$^{72+}$ ions can typically be accelerated to about 1 GeV/u and 400 MeV/u, respectively. The CSRe has a circumference of 128.8 m and a maximal magnetic rigidity of 9.4 Tm~\cite{Xia02,zwl2010}. 
\par
For ions stored in a storage ring, their revolution times, $t$, are related to the mass-to-charge ratios, $m/q$, via the following expression:
\begin{equation}
t=\frac{L}{c}\times\sqrt[]{1+\left( \frac{mc}{q}\right) ^2\times\frac{1}{(B\rho)^2}}~~~,
\label{eq1}
\end{equation}
where $L$ is the orbit length of the circulating ion, $c$ the speed of light, and $B\rho $ the magnetic rigidity. Since the ions within a certain acceptance of magnetic rigidity, $\Delta B\rho$, are all stored and circulate in the ring, their orbit lengths are not the same. In first order approximation, the relative time changes, $\Delta t/t$, (or equivalent relative revolution frequency changes $\Delta f/f$ ) are determined by~\cite{Rad00,Geis92,Fran08,Rad97}:
\begin{equation}
\frac{\Delta{t}}{t}
=-\frac{\Delta{f}}{f}=\alpha_p\times\frac{\Delta(m/q)}{(m/q)}+(\alpha_p\times\gamma^2-1)\times\frac{\Delta v}{v}~~, \label{eq2}
\end{equation}
where $\gamma$ is the relativistic Lorentz factor, $v$ the velocity of ions, and $\alpha_p $ the momentum compaction factor, which connects the relative change of the orbit length to the relative change of the magnetic rigidity of the circulating ions. $\alpha_p=-1/\gamma_t^2$ is nearly constant over the entire revolution frequency acceptance of the storage ring, where $\gamma_t$ is the so-called transition energy of the ring~\cite{Fran08}.
\par
From Eq.~(\ref{eq2}) it is clear that in order to determine the mass-to-charge ratios, $m/q$, of the stored ions, one needs to measure the revolution frequencies of the ions providing that the second term on the right hand side is negligibly small. This can be achieved by two techniques~\cite{Fran08}: The first one is related to the reduction of velocity spreads by applying beam cooling~\cite{Steck04}, and the revolution frequencies are then measured by applying the so-called Schottky Mass Spectrometry (SMS) technique~\cite{Litvi04,Litvi05,Litvi11,Bosch13}. The beam cooling requires several seconds, which sets a limit on the half-lives of nuclides that can be investigated using SMS. The second method is the IMS technique~\cite{Gels04,Haus00,Haus01,Sun08,Tu11}, which is based on a special beam-optics setting of the ring such that the injected ions have to fulfill the isochronous condition $\gamma=\gamma_t $. In such a case the velocity spreads of the stored ions are compensated by the lengths of closed orbits inside the ring and the revolution frequencies are a direct measure of the mass-to-charge ratios of the ions. The $B\rho$ acceptance of CSRe is $\pm 0.2$\% in the isochronous mode. 
\par
In the present experiment, a 467.91~MeV/u $^{58}$Ni$^{19+}$ primary beam of about $8\times 10^7$ particles per spill was fast-extracted and focused upon a $\sim $15~mm $^9$Be target placed in front of the RIBLL2. At this relativistic energy, the reaction products from projectile fragmentation of $^{58}$Ni were emerged from the target mainly as bare ions. They were then selected and analyzed~\cite{Geis92} by RIBLL2. A cocktail beam including the ions of interest was injected into the CSRe, which was tuned into the isochronous ion-optical mode~\cite{Haus00,Tu11} with the transition point at $\gamma_t=1.400$. The primary beam energy was selected according to the LISE++ simulations~\cite{Tar08} such that the $^{52}$Co$^{27+}$ ions had the most probable velocity with $\gamma=\gamma_t$ at the exit of the target. 
\par
Both RIBLL2 and CSRe were set to a fixed magnetic rigidity of $B\rho=5.8574$~Tm to allow for an optimal transmission of the $T_z=-1$ nuclides centered at $^{52}$Co. However, in the projectile fragmentation of $^{58}$Ni, fragments inevitably have broad momentum distributions of a few percent. All nuclides within the $B\rho$ acceptance $\pm 0.2$\% of the RIBLL2-CSRe system can be transmitted and stored in the CSRe. The $\gamma_t$ deviates from 1.400 at the edges of the CSRe aperture. A similar behavior has been reported in Ref.~\cite{Haus00} on the example of the ESR storage ring at GSI, Darmstadt. Therefore, we restricted the range of orbits by inserting a metal slit at a position with high dispersion (see Fig.\ref{Fig01}). The dispersion at the slit position  was estimated to be about 20 cm/\%. The slit opening was 60 mm corresponding to the momentum acceptance of $\Delta p/p\sim0.3\%$ in CSRe. As a result, the resolving power of the acquired spectra was considerably improved reaching $\sim 4\times 10^{5}$ (FWHM).
Typically, more than ten ions were
stored simultaneously in one injection.
The nuclides with well-known masses could be used as reference ions 
for mass calibration.
\par
The revolution times of the ions stored in CSRe were measured using a dedicated timing detector~\cite{Mei10} installed inside the CSRe aperture. 
It is equipped with a 19 $\mu$g/cm$^2$ carbon foil of 40 mm in diameter. Each time when an ion passed through the foil, secondary electrons were released from the foil and transmitted isochronously by perpendicularly arranged electric and magnetic fields to a Micro-Channel Plate (MCP) counter. The signals from the MCP were guided without amplification directly to a fast digital oscilloscope. In this experiment, we employed Tektronix DPO 71254 at a sampling rate of 50~GHz. The typical rise time of the signals was $0.25\sim 0.50$ ns~\cite{Mei10}.
\par 
The time resolution of the Time-of-Flight (ToF) detector was about 50 ps, and the detection efficiency varied from $\sim 20$\% to $\sim 70$\% depending on the charge and overall number of stored ions (See refs.~\cite{Tu11,Mei10} for more details). For each injection, a measurement time of  $300$~$\mu$s, triggered by the start pulse of the CSRe injection kicker, was set corresponding to about $500$ revolutions of the ions in CSRe. A total of 3840 injections were measured in the experiment. 
All ions that circulated for more than 100 $\mu s$ were considered in the analysis. This is different from previous analyses~\cite{Tu11,Ni53a,Yan13,Co51,TuPRL}, where the minimum time of 186 $\mu s$ within the measurement time of 200 $\mu s$ was required. Thus, the number of ions used in the analysis could be increased.
Following the procedures described in Ref.~\cite{Tu11}, we obtained the revolution time spectrum and made the particle identification. 

\subsection{Time shift correction} 

The particle identification in the revolution time spectrum can be obtained for a part of the experimental data acquired in a few hours~\cite{Tu11}. However, the peaks in the revolution time spectrum become broadened when several measurements are accumulated. This is caused by the instabilities of the magnetic fields of CSRe. Since the revolution times are used to determine mass values, the accuracies of the latter are dramatically affected by the drifts of the revolution times. Therefore, the time shifts due to the magnetic
field instabilities should be properly corrected. 
Previously one had to analyze a large number of sub-spectra obtained in a short period of measurement (typically several minutes)~\cite{Ni53a,Haus01}, or to use the correlation-matrix approach~\cite{Rad00,Sun08,Knobel}. In some cases, several reference ions were selected to construct the relative time spectra~\cite{Tu11}. Although, great improvements could be achieved, there were still some disadvantages as pointed out in Ref.~\cite{Shuai}. 
\par
Recently, we have developed a new method to minimize the deteriorations of revolution time spectra caused by the magnetic field instabilities. Our method is based on two assumptions: 
(1) The magnetic fields of the CSRe are stable during a short measurement time of $300~\mu$s, which allows us to make the time shift corrections on an injection-by-injection basis. (2) In a single injection, the time shifts from the ideal revolution times are constant for all stored ions. When the field changes are not too large, this assumption is valid for the ions in a limited revolution time range. Overall feasibility of the time shift correction can be checked by the succeeded mass calibration. Details of the correction method can be found in Ref.~\cite{Shuai}, and several key points are outlined here. 
\begin{figure} [b] 
	\includegraphics[angle=0,width=8.5cm]{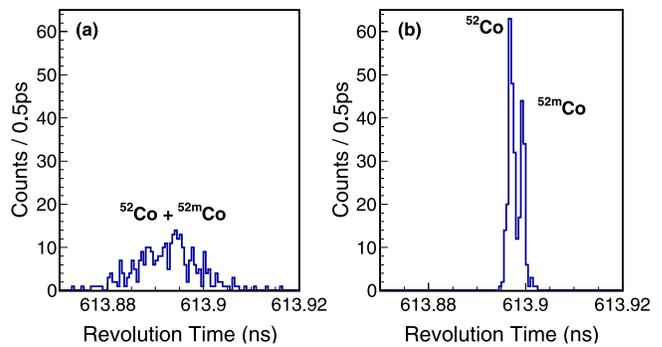}
	\caption{(Colour online) The revolution time spectra zoomed in around $^{52}$Co before (a) and after (b) time shift correction. The isomeric state ($T_{1/2}=104$~ms) at $E_x\sim390$ keV is clearly resolved from the ground state ($T_{1/2}=115$~ms) after the correction.
		\label{Fig02}}
\end{figure}

As a first step, we selected an ion species injected with high statistics and constructed a {\it relative} time spectrum~\cite{Tu11} from which the well-separated peaks can be identified. We used these peaks as {\it reference} ions, and calculated their mean revolution times and standard deviations (or equivalently the Root-Mean-Squares RMS): $\mu_i$ and $\sigma_i$ ~$(i=1,2,...,N_t)$ with $N_t$ being the total number of reference species. Peaks which were not well separated from others (for example $^{52g}$Co and $^{52m}$Co) have not been used as references. Then we went to the second step to make corrections on the injection-by-injection basis. For each individual injection, we selected $N_{ref}$ reference species (note that $N_{ref} \le N_t$) 
and calculated a weighted average time shift in this injection defined as:
\begin{equation}
\delta t
=\sum_{i=1}^{N_{ref}} \frac{1}{\sigma_{i}^{2}}\times (t_{i}-\mu_i) / \sum_{i=1}^{N_{ref}} \frac{1}{\sigma_{i}^{2}}.~~~\label{eq3}
\end{equation}
We subtracted this average shift from the initial revolution times $t_i$~$(i=1,2,...,N_s)$ for all the $N_s$ ions in this injection. We note that $N_{ref} \le N_s$, and only the injections with $N_{ref} \ge 2$ were used for the analysis. After all injections were analyzed in the same way, we had an {\it intermediate} time spectrum from which a new set of 
$\mu_i^{\prime}$ and $\sigma_i^{\prime}$ $(i=1,2,...,N_t)$ were obtained. Then, we replaced the \{$\mu_i$, $\sigma_i$\} with \{$\mu_i^{\prime}$, $\sigma_i^{\prime}$\} and made the correction from the above-mentioned second step $iteratively$ until \{$\mu_i^{\prime}$, $\sigma_i^{\prime}$\} were converged. The converged \{$\mu_i^{\prime}$, $\sigma_i^{\prime}$\} values were considered to be the final corrected revolution times with their corresponding deviations RMS. We note that an extra-term was added to $\sigma_i^{\prime}$ in each iteration so that over-corrections could be avoided (see Ref.~\cite{Shuai} for details). Fig.~\ref{Fig02} presents the revolution time spectrum zoomed in around $^{52}$Co before and after the correction. One sees that the overall time shift effect shown in Fig.~\ref{Fig02}(a) has been largely removed. Consequently, the resolving power of the spectrum is significantly improved after correction, and the isomeric state at $E_x\sim390$ keV is clearly resolved from its ground state.
\begin{figure*}   
	\begin{center}
	\includegraphics[angle=0,width=18cm]{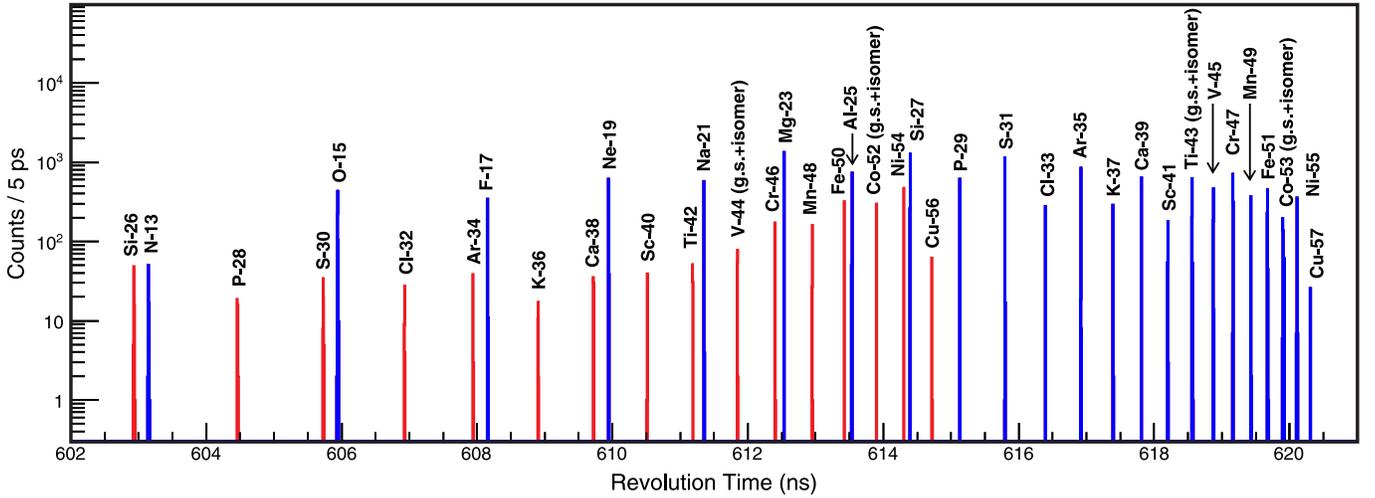}
	\caption{(Colour online) Part of the revolution time spectrum at a time window of $602~{\rm ns} \le t \le 621$~ns. The red and blue peaks represent the $T_z=-1$ and $-1/2$ nuclei, respectively. 
		\label{Fig03}}
	\end{center}
\end{figure*}
\begin{figure}
		\includegraphics[width=7.0cm]{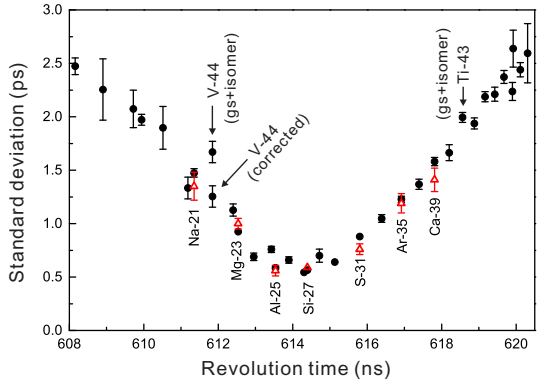}
		\caption{(Colour online) Plot of the calculated standard deviations or RMS of the revolution time peaks in Fig.~\ref{Fig03}. The black circles are the values deduced from the time-shift-correction method, while the red triangles are the values deduced directly from the two ions of same species stored simultaneously in the ring. Note that the latter are independent of the correction methods. The labels (gs+isomer) indicate that the RMS values are deduced from the mixed peaks of ground state and a low-lying isomer; the label (corrected) corresponds to the extrapolated RMS.  
			\label{Fig04}}
\end{figure}
\begin{figure}
	\begin{center}
		\includegraphics[width=8.0cm]{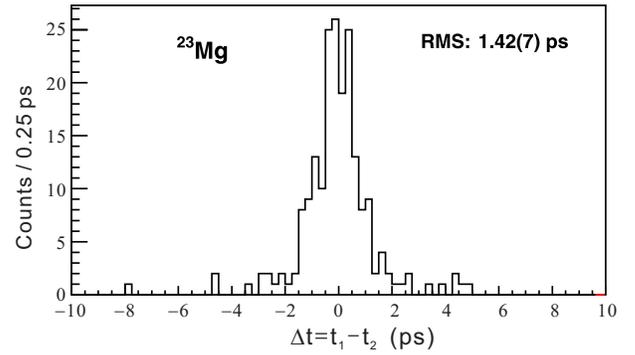}
		\caption{Distribution of the revolution time difference, $\Delta t=t_1-t_2$, obtained from injections where two $^{23}$Mg ions were stored simultaneously. The RMS divided by $\sqrt{2}$ is plotted in Fig.~\ref{Fig04}.  
			\label{Fig05}}
	\end{center}
\end{figure}

\par
Fig.~\ref{Fig03} illustrates a part of the corrected spectrum at a time window of $602~{\rm ns} \le t \le 621$~ns. The calculated standard deviations or the Root-Mean-Squares (RMS) for some of the revolution time peaks are shown in Fig.~\ref{Fig04}. The reached resolving power is up to $\sim 4\times 10^5$ FWHM for the nuclei in the isochronous region at about 614 ns.  
\par
In order to verify that the time spectrum is properly corrected and no over-correction occurred by the method itself,
we selected injections in which two ions of same species were present with revolution times $t_1$ and $t_2$. Their time difference, $\Delta t=t_1-t_2$, should be solely due to the momentum spreads and betatron oscillations of the ions in this injection. Analyzing all injections, we constructed a spectrum of $\Delta t$. An example for $^{23}$Mg is presented in Fig.~\ref{Fig05}. Since the magnetic fields keep stable in each individual measurement of $300~\mu$s, the width of the $\Delta t$ spectrum
should not be affected by the field instability. Such analyses have been applied to several ion species with high statistics, and the obtained $\sigma (\Delta t)/\sqrt{2}$ values are plotted in Fig.~\ref{Fig04}. One sees that the deduced $\sigma (\Delta t)/\sqrt{2}$ values are in good agreement with the calculated deviations from the correction method, indicating that the time shifts due to the magnetic field instabilities have been properly corrected. The mean revolution times and the calculated standard deviations obtained from the correction method are then used for mass calibrations and mass determinations.  

   
\begin{figure} 
	\begin{center}
		\includegraphics[width=9.cm]{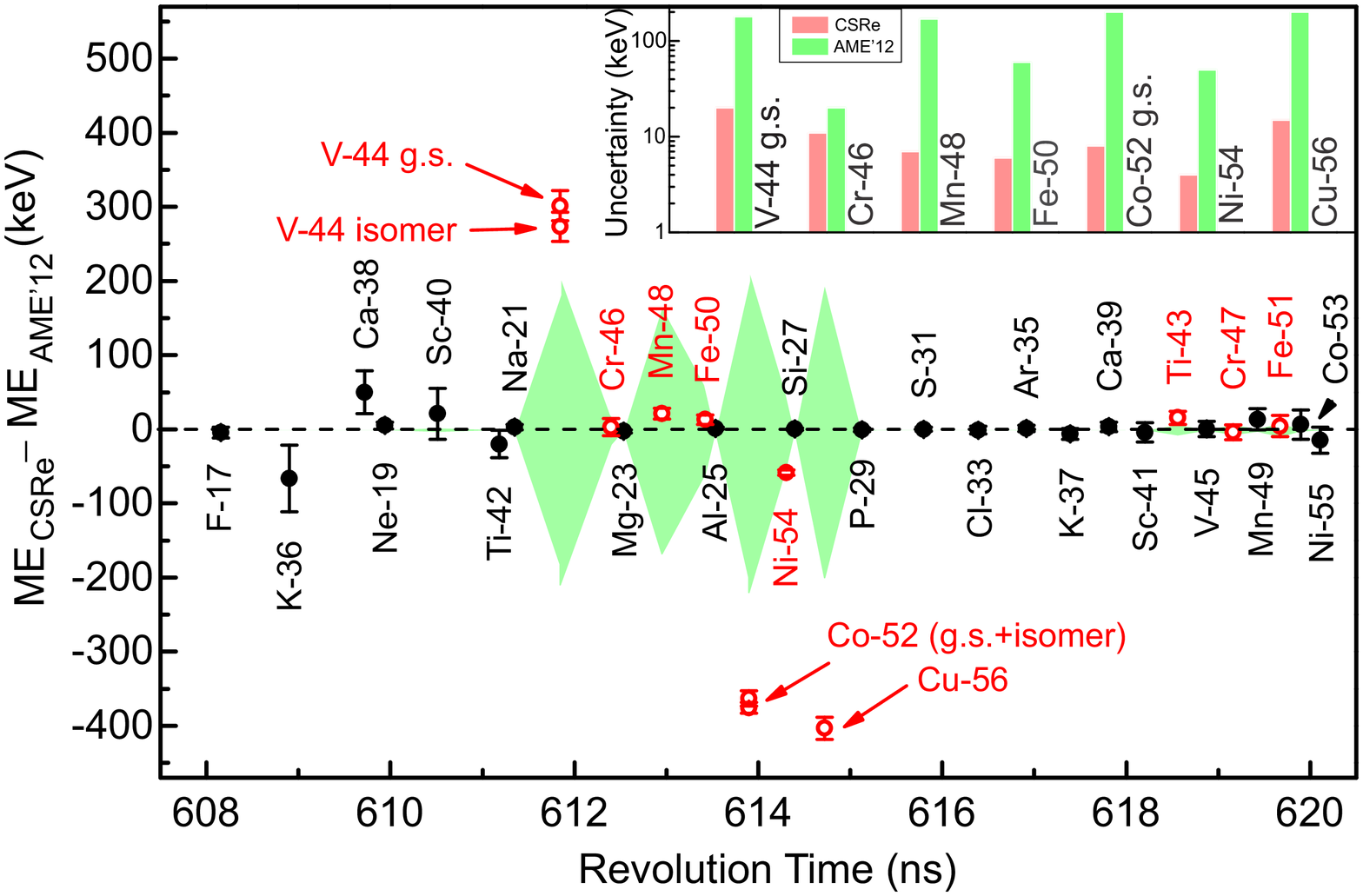}
		\caption{(Colour online)
			Differences between experimental $ME$ values determined in this work and those from the Atomic-Mass Evaluation AME$^{\prime}$12~\cite{AME2012} and Ref.~\cite{Kan14} for $^{45}$V and $^{49}$Mn. 
			The nuclides indicated by the filled symbols are used for calibration, which allowed for the mass determination of the nuclides indicated by open symbols. The value of each filled symbol corresponds to the re-determined mass using the other nuclides as calibrants except for itself.
			The green shadows indicate the $1\sigma$ error in AME$^{\prime}$12. The error bars in this figure represent the uncertainties in this measurement. The insert compares the uncertainties of our measurements with those in AME$^{\prime}$12~\cite{AME2012}.  
		}
		\label{Fig06}
	\end{center}
\end{figure}
\subsection{Mass determination} 
Two series of nuclides with $T_z=-1/2$ and $T_z=-1$ were stored in the CSRe and their revolution times have been measured in the experiment (see Fig.~\ref{Fig03}).  
Most of the masses of these nuclides are known with high precision,  
therefore we have used nuclides with experimental mass uncertainties of less than 5~keV 
to make the mass calibration according to:  
\begin{equation}
m/q(t)=a_o+a_1\times{t}+a_2\times{t^2}+a_3\times{t^3}~, \label{eq4}
\end{equation}
where $a_0$, $a_1$, $a_2$, and $a_3$ are free parameters. Since both literature
mass values and measured mean revolution times, $t's$, have uncertainties, the weights of the fitted
points have been taken as a quadratic combination of both errors as 
\begin{equation}
\sigma^2=\sigma^2_{m/q}+(a_1+2a_2\times{t}+3a_3\times{t^2})^2\times \sigma^2_t.
 \label{eq5}
\end{equation}
The least-squares fit has been done iteratively in the same way as
described, e.g., in Ref.~\cite{Matos}. In brief, at each iteration
the new fit parameters $a_1$, $a_2$, and $a_3$ are used to readjust the
weights $\sigma^2$ in Eq.~(\ref{eq5}).
The fitting has been performed with different starting values for
$a_1$, $a_2$, and $a_3$. In all cases the procedure converged to the
identical parameter values.
\par
Note that the ions stored in CSRe are fully stripped, therefore the atomic masses $M(A,Z)$ given in~\cite{AME2012} have been transformed into nuclear masses $m(A,Z)$ in the mass calibration of  Eq.~(\ref{eq4}) according to~\cite{Lunney03}:
\begin{equation}
m(A,Z)=M(A,Z)-Z\times m_e+B_e(Z),
\label{eq6}
\end{equation}
where $m_e$ is the mass of electron, and the total binding energy of all electrons, $B_e(Z)$, is estimated~\cite{Lunney03} by
\begin{equation}
B_e(Z)=14.4381\times Z^{2.39}+1.55468\times 10^{-6}\times Z^{5.35}~{\rm eV,}
\label{eq7}
\end{equation} 
which provides an RMS error over the entire range of tabulated masses of 150 eV. 
Eq.~(\ref{eq4}) with the parameters $a_0$, $a_1$, $a_2$, and $a_3$ obtained from the fitting procedure was then used to determine the masses of $T_z=-1,-1/2$ nuclei of interest via interpolation. Finally the determined nuclear masses are transformed back into the atomic masses using Eqs.~(\ref{eq6}) and (\ref{eq7}).
\par 
In order to check the reliability of mass calibration and examine the possible systematic error of our approach, we have re-determined the Mass Excess $(ME)$ of each of the $N_c$ calibrant nuclides using the other $N_c-1$ ones for mass calibration. The differences between the re-determined $ME$-values and the literature ones~\cite{AME2012} are compared in Fig.~\ref{Fig06}. The normalized $\chi_{n}$ is calculated according to 
\begin{equation} \label{eq8}
\chi_{n}=\sqrt{\frac{1}{n_f}\sum_{i=1}^{N_c}\frac{(ME_{{\rm CSRe},i}-ME_{{\rm AME},i})^2}{\sigma_{{\rm CSRe},i}^2+\sigma_{{\rm AME},i}^2}},
\end{equation}
where $n_f=N_c$ is the number of degrees of freedom, $ME_{{\rm CSRe},i}$ the mass excess measured in CSRe, and $ME_{{\rm AME},i}$ the corresponding tabulated mass excess from the 2012 Atomic-Mass Evaluation. Since some of our data were collected as private communication already in the Atomic-Mass Evaluation AME$^{\prime}$2016~\cite{AME2016}, we have to compare our results to the ones published in AME$^{\prime}$12.
If the calculated $\chi_{n}$ is within the expected range of $\chi_{n}=1\pm 1/\sqrt{2n_f}$ at $1\sigma$ confidence level, systematic errors would not be considered. If $\chi_{n}$ is outside the $1\sigma$ limits, a systematic error, $\sigma_{sys}$, has to be added to the final mass values. This systematic error is obtained by introducing $\sigma_{sys}^2$ in the denominator of Eq.~(\ref{eq8}) with the requirement of $\chi_{n}=1$~\cite{Matos}. 

\section{Experimental Results}
The upper $fp$-shell nuclei of interest were tuned into the isochronous region of $611~{\rm ns} \le T \le 615$~ns, and their revolution times were measured with high precision (see Figs.~\ref{Fig03} and~\ref{Fig04}). We used $N_c=21$ nuclides in the time window of $608~{\rm ns} \le T \le 621$~ns to calibrate the revolution time spectrum. The masses of these calibrants have been redetermined using the method described above yielding $\chi_{n}=1.09$. This value is within the expected range of $\chi_{n}=1\pm0.15$ at $1\sigma$ confidence level, indicating that no additional systematic errors have to be considered. The excellent agreement between the redetermined masses and the literature ones (see Fig.~\ref{Fig06}) confirms again the validity of the time shift corrections. For the well-separated peaks in Fig.~\ref{Fig03}, we used the calculated mean revolution times and corresponding RMS to determine the nuclear masses. When two peaks were not completely separated, the spectrum-decomposition or maximum likelihood methods have been used to extract the reliable mean revolution times and their corresponding deviations for the mass determinations.
 
\subsection{Masses of $^{46}$Cr, $^{48}$Mn, $^{50}$Fe, $^{54}$Ni, and $^{56}$Cu} 
The masses of $^{46}$Cr, $^{50}$Fe, and $^{54}$Ni were measured about 40 years ago either through a reaction excitation function~\cite{Zi72} or through the $Q$ values of ($^4$He,$^8$He) reactions~\cite{Tr77}. The mass of $^{54}$Ni was addressed at the JYFLTRAP
Penning-trap mass spectrometer, but no result could be obtained due to a very low production yield~\cite{Kan10}. The mass of $^{48}$Mn was measured for the first time in the
storage ring ESR of GSI using the IMS technique~\cite{Stad04}. Two settings of the magnetic fields were employed leading to an averaged mass excess with the evaluated uncertainty of  170~keV~\cite{AME2012}. No experimental mass for $^{56}$Cu had been reported prior to the present experiment. 
\par
The $ME$ values of $^{46}$Cr, $^{48}$Mn, $^{50}$Fe, $^{54}$Ni, and $^{56}$Cu are obtained in this work and are presented in Table~\ref{table01} together with the values from AME$^{\prime}$12~\cite{AME2012}. The differences are given in the last column of Table~\ref{table01} and are shown in Fig.~\ref{Fig06}. 
Our measured atomic masses are in good agreement with the values in AME$^{\prime}12$~\cite{AME2012}, but with much improved precision. In particular, the newly determined masses of $^{48}$Mn, $^{50}$Fe, and $^{54}$Ni are one order of magnitude more precise than the adopted values~\cite{AME2012}, and a relative mass precision of $\sim 1\times 10^{-7}$ has been achieved for $^{54}$Ni. 
\par
Our measurement yields $ME(^{48}$Mn)$=-29299(7)$~keV, which is in excellent agreement with the value of Ref.~\cite{Stad04} in their first magnetic field setting but 17 times more precise.
This result can independently be verified using the recent data from the $\beta$ decay studies of $^{48}$Fe~\cite{Orrigo16}. In that experiment, both $\beta$-delayed protons and $\beta$-delayed $\gamma^{\prime}$s of $^{48}$Fe have been measured and assigned as being due to the de-excitation from the $T=2$ Isobaric Analog State (IAS) in the $T_z=-1$ nucleus $^{48}$Mn. Therefore the ground-state $ME$ value of $^{48}$Mn can be obtained through:
\begin{equation} \label{eq10}
\begin{split}
& ME(^{48}{\rm Mn})=ME(^{47}{\rm Cr})+ ME(^1{\rm H})+E_p^{\rm c.m}-E_{\gamma}^{\rm sum}\\
& ~~~~~~~~=[-34561(7)+7289+1018(10)-3036.5(1.5)]\\
& ~~~~~~~~=-29291(12)~~{ \rm keV},
\end{split}
\end{equation}
with $E_p^{\rm c.m}$ being the center-of-mass proton decay energy, and $E_{\gamma}^{\rm sum}$ the summed energy of three sequential $\gamma$ transitions from the IAS in $^{48}$Mn. The deduced $ME$ value for $^{48}$Mn is in excellent agreement with our result.
 
\begin{table}
	\caption{Experimental $ME$ values obtained in this work and
		values from the Atomic-Mass Evaluation AME$^{\prime}$12~\cite{AME2012} and Ref.~\cite{Kan14} for $^{45}$V and $^{49}$Mn. The deviations $\delta ME={ME}_{\rm CSRe}-{ME}_{\rm AME'12}$ are given in the last column. Also listed are the numbers of identified ions $N$, standard deviations, $\sigma_t$,  and
		FWHM values of the revolution time peaks (see Fig.~\ref{Fig02}) converted in keV through
		mass calibration.}
	\begin{tabular}{ccccccc} \hline\hline
		Atom &  $N$ & $\sigma_t$ & FWHM & $ME_{\rm CSRe}$   & $ME_{\rm AME'12}$ &  $\delta ME$ \\
		&       & (ps)       & (keV)& (keV)             &  (keV)            &(keV) \\[1mm]
		\hline
		$^{44g}$V & 64 & 1.25 & 382 & $-23827(20) $  & $-24120(180)     $ & $294(180)      $ \\
		$^{44m}$V & 75 & 1.25 &382& $ -23541(19)$ & $ -23850(210)^{\#} $ & $ 309(210)^{\#} $ \\
		$^{46}$Cr & 195 &1.13 & 373 &$-29471(11) $  & $-29474(20)      $ & $ 3(23)      $ \\
		$^{48}$Mn & 198 & 0.68 & 242 &$-29299(7)  $  & $-29320(170)     $ & $21(170)      $ \\
		$^{50}$Fe & 342 & 0.76 & 277 &$-34477(6)  $  & $-34490(60)      $ & $13(60)      $ \\
		$^{52g}$Co & 194 & 0.66 & 246 &$ -34361(8)$  & $-33990(200)^{\#}$ & $-371(200)^{\#} $ \\
		$^{52m}$Co& 129 & 0.75 & 277 &$ -33974(10) $  & $-33610(220)^{\#}$ & $-364(220)^{\#} $ \\
		$^{54}$Ni & 688 & 0.54 & 226 & $-39278(4)  $  & $-39220(50)      $ & $-58(50)     $ \\
		$^{56}$Cu & 64  & 0.70 & 276 &$-38643(15) $  & $-38240(200)^{\#} $ & $-403(200)^{\#} $ \\
		$^{43}$Ti & 920 & 1.99 & 631 &$ -29306(9)^{**}$    & $-29321(7)$ & $ 15(11) $ \\
		$^{45}$V  & 687 & 1.94 & 651 &$ -31885(10)$    & $-31885.3(9)^{*}$ & $ 0(10) $ \\
		$^{47}$Cr & 1083 & 2.19 & 791 &$ -34565(10)$    & $-34561(7)       $ & $-4(12) $ \\
		$^{49}$Mn & 561 & 2.21 & 816 & $-37607(14)$    & $-37620.3(24)^{*} $ & $13(14)     $ \\
		$^{51}$Fe & 760 & 2.37 & 932 &$-40198(14) $    & $-40202(9)  $ & $4(17) $ \\
		
		\hline \hline
	\end{tabular}
	\flushleft{$^{\#}$Extrapolated values in AME$^{\prime}$12~\cite{AME2012}. $^{**}$Mixed with the $E_x=313(1)$~keV low-lying isomer~\cite{AME2012}.  $^{*}$Values from latest measurements in Ref.~\cite{Kan14}.} 
	{\normalsize } \label{table01}
\end{table}
\begin{figure} [b]
	\begin{center}
		\includegraphics[width=7 cm]{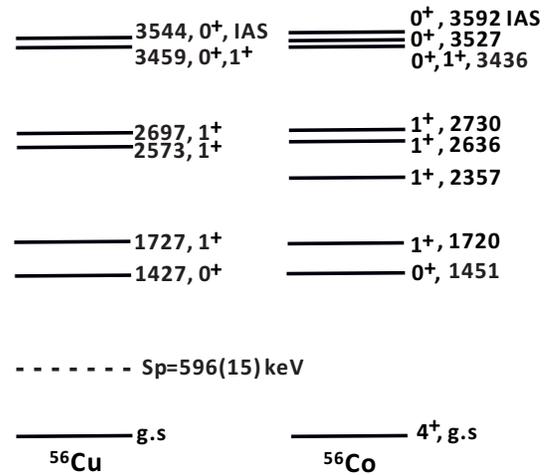}
		\caption{ Excited low-spin states in the mirror nuclei $^{56}$Cu and $^{56}$Co. The excited states in $^{56}$Cu are from Ref.~\cite{56Zn} but shifted by 36 keV upward according to our ground-state mass of $^{56}$Cu. The data for $^{56}$Co are from Ref.~\cite{56Co}.     
			\label{Fig07}}
	\end{center}
\end{figure}

The mass excess of $^{56}$Cu is measured to be $ME=-38643(15)$~keV, which has been collected already in AME$^{\prime}$2016~\cite{AME2016}.
This $ME$ is consistent with the AME$^{\prime}03$ value~\cite{AME03} but differs by 403~keV from the extrapolated one in AME$^{\prime}12$~\cite{AME2012}. Our mass excess yields the proton separation energy of $^{56}$Cu to be $S_p=596(15)$~keV, which is in excellent agreement with the predicted value, $S_p=620$~keV, using the Coulomb displacement energy calculations~\cite{Kanebo13}. Using this proton separation energy, the excited states in $^{56}$Cu identified from the $\beta$-delayed proton decays of $^{56}$Zn~\cite{56Zn} should be shifted upward by 36~keV, leading to an amazing mirror symmetry of excited levels between $^{56}$Cu~\cite{56Zn} and $^{56}$Co~\cite{56Co} (see Fig.~\ref{Fig07}). Finally, it is worth noting that our mass excess of $^{56}$Cu agrees with $ME=-38626.7(7.1)$~keV which is reported in Ref.~\cite{Valv2018} during the review process of this manuscript.  

\subsection{Masses of $^{52g}$Co and $^{52m}$Co} 
Experimental masses did not exist for the ground state of $^{52}$Co and its $(2^+)$ isomer~\cite{AME2012}. Considering mirror symmetry of nuclear levels~\cite{Tu16}, and using the data of $\beta$-delayed protons and $\beta$-delayed $\gamma ^{\prime}$s of $^{52}$Ni, the mass excesses of the ground and isomeric states had been addressed in Refs.~~\cite{Dossat,Orrigo16}.
During the preparation of this manuscript, the $ME$-values of $^{52g}$Co and $^{52m}$Co from JYFLTRAP mass spectrometry have been reported~\cite{Nest2017}, see Table~\ref{table02}. 
\par
\begin{figure} [b]
	\begin{center}
		\includegraphics[width=8.5 cm]{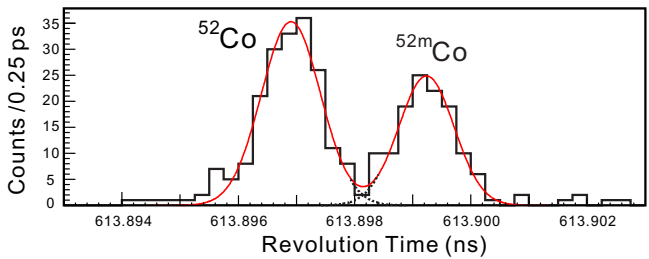}
		\caption{(Colour online) The revolution time spectrum zoomed in around $^{52}$Co. The red solid line represents the fitted result using two Gaussian distributions shown as black dotted lines. 
			\label{Fig08}}
	\end{center}
\end{figure}

In our work, the $^{52g}$Co and $^{52m}$Co were very well resolved, see Fig.~\ref{Fig08}. 
However, the two peaks are not exactly symmetric, one observes few counts at larger revolution times outside three standard deviations. 
Therefore we employed different ways to find their centroids and the corresponding uncertainties, namely the double-Gaussian chi-squares fitting, unbinned maximum likelihood estimation, and unbinned Bayesian analysis. We also decomposed the two peaks using two-Gaussian functions and then calculated the mean revolution times and the corresponding RMS. The mass excesses of $^{52g}$Co and $^{52m}$Co obtained from the four approaches are presented in Table~\ref{table02}, together with the literature values in Refs.~\cite{Nest2017,AME2012} for comparison.
\par
Our $ME$ values from the last three approaches are in excellent agreement with each other. The binned double-Gaussian chi-squares fitting yields $ME$ values which deviate a little from the former ones. This is due to the $outlying$ counts outside 3 standard deviations of the peaks. In Table~\ref{table01} and Ref.~\cite{Xu16}, the $ME$ values from the unbinned maximum likelihood estimation were adopted. We note that our $ME$ value of $^{52m}$Co is consistent within $1\sigma$ with the one in Ref.~\cite{Nest2017}, whereas our $ME(^{52g}\rm Co)$ is 29 keV lower ($2.8\sigma$) than the one in \cite{Nest2017}. The new proton separation energy of $^{52g}$Co $S_p=1447(12)$~keV agrees well with the predicted value of $S_p=1.54$~MeV in Ref.~\cite{Kanebo13}.  
\par
The determined mass excesses for the ground state of $^{52}$Co and its $(2^+)$ isomer are $\sim 370$~keV more bound than the extrapolated values in AME$^{\prime}12$~\cite{AME2012} while they are $\sim 130$~keV less bound than the value deduced in Ref.~\cite{Orrigo16}. These energy differences have significant consequences in the understanding of the $\beta$-decay properties of $^{52}$Ni that has been discussed in our previous publication~\cite{Xu16}. The excitation energy of the isomer is thus determined to be 387(13)~keV, which is very close to $E_x=378(1)$~keV of the 2$^+$ isomer in its mirror nucleus $^{52}$Mn~\cite{52Mn73}. The impact of this result on the IMME test will be discussed in section IV.

\begin{table} 
	\caption{Experimental $ME$ values of $^{52g}$Co and $^{52m}$Co obtained in this work using different analysis methods: (1) double-Gaussian chi-squares fitting, (2) unbinned maximum likelihood estimation, (3) unbinned Bayesian analysis, and (4) peak decomposition using two-Gaussian functions. $ME$ values from Refs.~\cite{Nest2017} and~\cite{AME2012} are also given for comparison. The deviations $\delta ME={ME}_{\rm CSRe}-{ME}_{\rm AME'12}$ are given in the 4th and 5th columns. Also listed are the excitation energy, $E_x(2^+)$, of $^{52m}$Co.}
	\begin{tabular}{cccccc} \hline\hline
		Method &  $ME(^{52g}$Co) & $ME(^{52m}$Co) & $\delta ME_{gs}$ & $\delta ME_{is}$ & $E_x(2^+)$  \\
		&  (keV)     & (keV)       & (keV)& (keV)  & (keV)  \\[1mm]
		\hline
		(1) & -34355(7) & -33997(8) & -365(7) &-387(8)& 358(11)   \\
		(2) & -34361(8) & -33974(10) & -371(8) &-364(10)&  387(13)   \\
		(3) & -34360(9) &-33976(13) & -370(9) &-366(13)&  384(16)  \\
		(4) & -34366(8) & -33975(11) & -376(8) &-365(11)&  391(14)  \\
		\cite{Nest2017}& -34331.6(66) & -33958(11) & -341.6(66) &-348(11)& 374(13)  \\
		\cite{AME2012}  & -33990(200) & -33610(220) & -- &--& 380(100) \\
		
		\hline \hline
	\end{tabular}
	{\normalsize } \label{table02}
\end{table}
 
\subsection{Masses of $^{44g}$V and $^{44m}$V} 
Prior to this experiment, the mass of $^{44}$V had been measured only in the storage ring ESR of GSI using the IMS technique~\cite{Stad04}. Due to the unresolved isomer ($T_{1/2}=150$~ms) at a predicted excitation energy of $E_x\sim270$~keV~\cite{AME2012}, the mass excess of $^{44}$V was reported with an asymmetric uncertainty as $ME=-23980^{+80}_{-380}$~keV, which has been evaluated and recommended to be $ME=-24120(180)$~keV in AME$^{\prime}$12~\cite{AME2012}.  
\par
\begin{figure} [t]
 	\begin{center}
 		\includegraphics[width=8.5 cm]{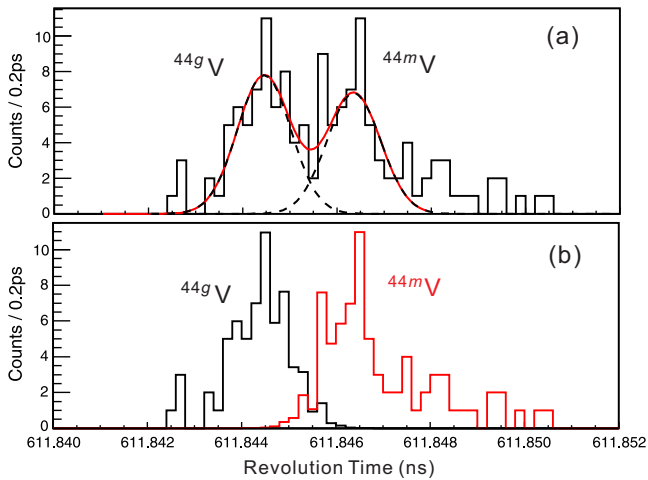}
 		\caption{(Colour online) (a) The revolution time spectrum zoomed in around $^{44}$V. The red solid line represents the fitted result using two Gaussian distributions shown as black dashed lines. (b) Decomposed time spectra of the ground-state $^{44}$V and its $(6^+)$ isomer according to the fitted Gaussian distributions. 
 		\label{Fig09}}
 	\end{center}
 \end{figure}

In the present experiment, the $^{44}$V$^{23+}$ ions at both ground and isomeric states were stored in CSRe, and their revolution times were measured with high precision. The expanded revolution time spectrum centered at $^{44}$V is presented in Fig.~\ref{Fig09}. Although $^{44}$V was not in the isochronous condition (see Fig.~\ref{Fig04}), the two peaks can still be resolved.
\par            
Since the statistics for $^{44}$V$^{23+}$ ions is lower in this experiment, we have tried to extract the centroids and corresponding uncertainties using two approaches, namely the double-Gaussian chi-squares fitting (see Fig.~\ref{Fig09}(a)), and the spectrum decomposition using two-Gaussian functions (see Fig.~\ref{Fig09}(b)). The latter approach yields a smaller uncertainty for the centroid of $^{44g}$V ($\sigma_t=0.70$ ps), which obviously deviates from the general trend as shown in Fig.~\ref{Fig04}. Therefore $\sigma_t=1.25$ ps obtained from interpolation in Fig.~\ref{Fig04} was used in error estimation. The $ME$ values of $^{44g}$V and $^{44m}$V determined from the two analysis methods  
are consistent with each other within $2\sigma$. The averaged values are adopted here and given in Table~\ref{table01}.
\par
The mass excesses determined in this work are $ME(^{44g}$V$)=-23827(20)$~keV and $ME(^{44m}$V$)=-23541(19)$~keV, leading to an excitation energy $E_x=286(28)$~keV for the $(6^+)$ isomer. On the one hand, this excitation energy is close to the value of $E_x=271$~keV in its mirror nucleus $^{44}$Sc~\cite{AME2012}, and is in agreement with $E_x=266$~keV deduced from calculated Mirror-Energy Difference (MED) of $-5$~keV between $^{44}$V and $^{44}$Sc~\cite{Tay11}. On the other hand, using the ground state mass obtained in this work, the excitation energy of the $T=2$ IAS of $^{44}$Cr in $^{44}$V is deduced to be 2703(24)~keV rather than 2990(180)~keV in~\cite{AME2012}, which is comparable with $E_x=2778(3)$~keV of the $T=2$ analog state in $^{44}$Sc~\cite{AME2012}. Furthermore, using the present mass excess, the $S_p(^{45}{\rm Cr})=2972(45)$~keV can be determined rather than $S_p(^{45}{\rm Cr})=2690(190)$~keV in Ref.~\cite{AME2012}. The larger $S_p(^{45}{\rm Cr})$ value makes the Ca-Sc circling in the $rp$-process more unlikely as we have pointed out in Ref.~\cite{Yan13}. 

\subsection{Masses of $T_z=-1/2$ nuclei $^{45}$V, $^{47}$Cr, $^{49}$Mn, and $^{51}$Fe} 

The masses of the $T_z=-1/2$ {\it fp}-shell nuclides $^{45}$V, $^{47}$Cr, $^{49}$Mn, and $^{51}$Fe have been determined previously at Michigan State University~\cite{Pro72,Mue75}. They were also measured in the CSRe using the IMS technique via the projectile fragmentation of $^{78}$Kr~\cite{Tu11}. Recently, high-precision $Q_{EC}$-value measurements with the double Penning-trap mass spectrometer JYFLTRAP have been performed for $^{45}$V and $^{49}$Mn~\cite{Kan14}. A 22-keV (2$\sigma$) deviation has been found for $^{49}$Mn compared to the value of Ref.~\cite{Tu11}. The authors called for further measurements of atomic masses obtained in CSRe to confirm or disprove the observed  breakdown of the quadratic form of the isobaric multiplet mass equation in the $fp$ shell~\cite{Ni53a}. 
\par
In the present experiment, the ion$^{\prime}$s orbits are restricted by inserting a slit in the CSRe aperture, leading to a considerably improved resolving power in the time spectra. Although the $^{45}$V, $^{47}$Cr, $^{49}$Mn, and $^{51}$Fe ions were not in the best isochronous region, their masses could still be obtained with good precision. Following the procedures mentioned above, their mass excesses have been re-determined. The results are presented in Table~\ref{table01} and compared with literature values in Fig.~\ref{Fig06}. 
Our re-determined $ME(^{49}{\rm Mn})=-37607(14)$~keV is 35(17)~keV larger than the previous CSRe value of $-37642(11)$ keV, but it is in good agreement with the JYFLTRAP result of $ME(^{49}{\rm Mn})=-37620.3(24)$~keV. For the other three nuclides, the $ME$ values obtained in both CSRe experiments are consistent with each other and the mass of $^{45}$V agrees well with that from the JYFLTRAP experiment~\cite{Kan14}. It is worth noting that the consistent $ME$-value of $^{51}$Fe from two independent measurements provides solid ground to deduce the energies of excited states in $^{52}$Co via the $\beta$-delayed proton emissions of $^{52}$Ni~\cite{Xu16}. 
\par
Finally, we note that the re-determined $ME$ value for $^{43}$Ti in this experiment, assuming a single peak, is 15~keV larger than the AME$^{\prime}12$ value. This could be understood as due to the contamination of a low-lying isomer at $E_x=313(1)$~keV~\cite{AME2012}. Although the half-life of this isomer is short ($T_{1/2}=11.9~\mu$s), it may live much longer than the neutral atoms when fully stripped~\cite{Yuri2003,Sun2010,Zeng2017}. In fact, the broadened peak of $^{43}$Ti had been observed in the IMS experiments in the ESR of GSI~\cite{Stad04} and the CSRe of IMP~\cite{Tu11,TuPRL}. 

\section{Discussion }

\subsection{$A$ dependence of vector and tensor Coulomb energies in $T_z=-1$, $fp$-shell nuclei} 
\par
The masses of the members of an isobaric multiplet are given by~\cite{Jan66} 
\begin{equation}
M(A,T,T_z)=M_0(A,T)+E_c(A,T,T_z)+T_z\times \Delta m,
\label{eq10}
\end{equation}
where $\Delta m=782$~keV is the neutron-hydrogen mass difference and $E_c$ the Coulomb energy. Present measurements complete the ground-state masses of the $T_z=-1$, $fp$-shell nuclei up to the heaviest one $^{58}$Zn. In addition, excitation energies for the $T=1$ IAS$^{\prime}$s in the self-conjugate nuclei have already been known with high precisions~\cite{AME2012}. These new data together with the older ground-state masses for the 
$T_z=+1$ nuclei~\cite{AME2012} can be used to derive Coulomb energy differences, $\Delta E_c(A,T,T_z|T_z^{\prime})$, defined as
\begin{equation} \label{eq11}
\begin{split}
&\Delta E_c(A,T,T_z|T_z^{\prime})\equiv E_c(A,T,T_z)-E_c(A,T,T_z^{\prime}) \\
& =M(A,T,T_z)-M(A,T,T_z^{\prime})-(T_z-T_z^{\prime})\times \Delta m.
\end{split}
\end{equation}
The Coulomb energy differences are related to the so-called vector and
tensor Coulomb energies~\cite{Jan66}, $E_c^{(1)}(A,T)$ and $E_c^{(2)}(A,T)$, respectively. For the $T=1$ isobaric triplets, these energies can be obtained through the following expressions 
\begin{gather}
	E_c^{(1)}(A,1)=[\Delta E_c(A,1,-1|0)+\Delta E_c(A,1,0|+1)]/2,\label{eq12}\\
	E_c^{(2)}(A,1)=[\Delta E_c(A,1,-1|0)-\Delta E_c(A,1,0|+1)]/6.\label{eq13}
\end{gather}
The vector and tensor Coulomb energies are independent of the isospin projection, $T_z$, and are related to the coefficients in the Isobaric Multiplet Mass Equation    
\begin{equation}
M(A,T,T_z)=a(A,T)+b(A,T)T_z+c(A,T)T^2_z,
\label{eq14}
\end{equation}
with 
\begin{gather}
	b(A,T)=\Delta m-E_c^{(1)}(A,1),\label{eq15}\\
	c(A,T)=3\times E_c^{(2)}(A,1).\label{eq16}
\end{gather}

\begin{figure}
	\begin{center}
		\includegraphics[width=8.5 cm]{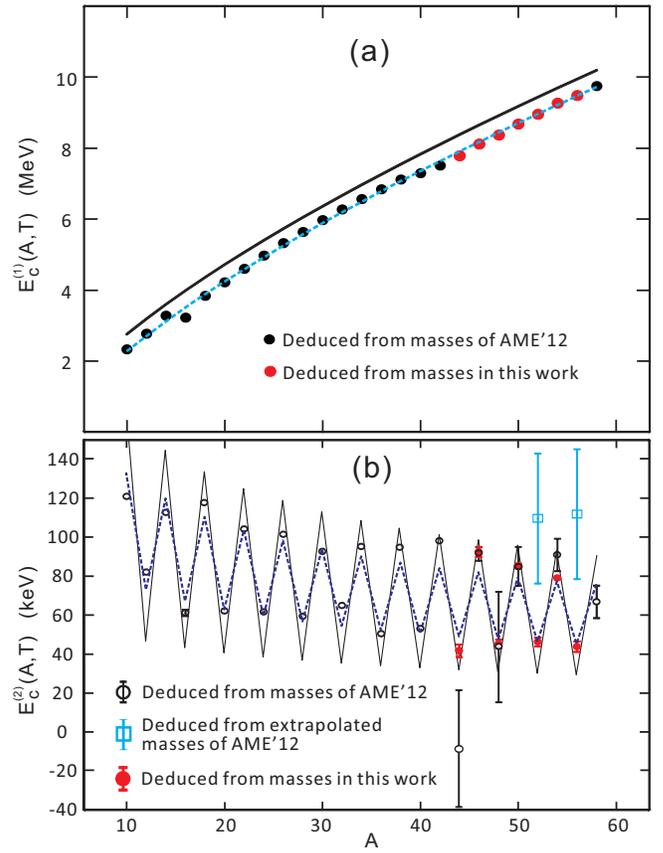}
		\caption{(Colour online) Plots of (a) vector and (b) tensor Coulomb energies as a function of $A$. The solid lines represent the calculated results based on Eq.~(\ref{eq17}). The dashed line in the upper panel is from the same calculations by replacing the coefficient 0.46 of the second term in Eq.~(\ref{eq17}) by 0.828. The dashed line in the lower panel is from the same calculations by replacing the coefficient 0.15 of the third term in Eq.~(\ref{eq17}) by 0.075.    
			\label{Fig10}}
	\end{center}
\end{figure}

Studies on $E_c^{(1)}(A,T)$ and $E_c^{(2)}(A,T)$ (or equivalently on $b$ and $c$ coefficients) and their $A$ and $T$ dependence have since long been an important research subject~\cite{Jan66,Nolen69,Shlomo78,Ben07,Lam13,Mac14}. A detailed knowledge of the $A$ dependence of $E_c^{(2)}(A,T)$ can yield information on the charge-violating nuclear forces~\cite{Jan66,Lam13}. Such an $A$ dependence of $E_c^{(1)}(A,T)$ and $E_c^{(2)}(A,T)$ has been extracted from the experimental masses according to Eqs.~(\ref{eq11})$-$(\ref{eq13}) and plotted in Fig.~\ref{Fig10}. It is seen that the general features have been well established by adding our mass data, especially, an oscillation in $E_c^{(2)}(A,T)$ with smoothly changing amplitude is clearly evidenced up to $A=58$. The previous experimental mass for $^{44}$V, and the extrapolated masses for $^{52}$Co and $^{56}$Cu~\cite{AME2012} result in significant deviations from the general trend of  $E_c^{(2)}(A,T)$ versus $A$ (see Fig.~\ref{Fig10}(b)). 
The oscillatory structure in the tensor Coulomb energy has been observed in the limited $sd$-shell nuclei and attributed to the Coulomb pairing effects~\cite{Jan66} as well as to the non-negligible contribution of isospin non-conserving forces of nuclear origin~\cite{Lam13}. However, it has been unknown if such oscillation phenomenon persists when entering into the $fp$-shell. Our mass data measured in this work demonstrate definitely the persistence of oscillatory structure in $fp$-shell, and provide a test ground in $fp$-shell for investigating the effects of isospin symmetry breaking.      
\par
From Eq.~(\ref{eq11}) one sees that the Coulomb energy difference, $\Delta E_c$, can be calculated. We have used the
semi-classical approach~\cite{Nolen69,Shlomo78}
\begin{equation}
E_c=\{0.6Z^2-0.46Z^{4/3}-0.15[1-(-1)^Z]\} \times \frac{e^2}{r_0A^{1/3}},
\label{eq17}
\end{equation}
to calculate the $A$ dependence of $E_c^{(1)}(A,T)$ and $E_c^{(2)}(A,T)$. In this approach, the nucleus is assumed to be a uniformly charged sphere. Also the exchange and Coulomb pairing energies are included (second and third terms). 
 The calculated results with $r_0=1.25$ are plotted in Fig.~\ref{Fig10} as solid lines. One sees that the general trend of $E_c^{(1)}(A,T)$ and the oscillation pattern of $E_c^{(2)}(A,T)$ can be reproduced, indicating that the oscillation in $E_c^{(2)}(A,T)$ persists in the heavier mass region. However, the calculated $E_c^{(1)}(A,T)$ values are systematically larger ($\sim 500$~keV) than those deduced from the experimental masses (see Fig.~\ref{Fig10}(a)), 
 and the oscillation amplitude is over-predicted. Interestingly, the over-estimation of $E_c^{(1)}(A,T)$ can be removed by multiplying the exchange term of Eq.~(\ref{eq17}) by 1.8 (see the dashed line in Fig.~\ref{Fig10}(a)). This modification has been used in the calculations of symmetry energy coefficients of finite nuclei~\cite{Tian14}. Similarly the theoretical oscillation amplitude in Fig.~\ref{Fig10}(b) can be reduced by $artificially$ multiplying the pairing term of Eq.~(\ref{eq17}) by 0.5. The reasons for this are not yet understood and require further investigations.  
\par


 \subsection{Test of nuclear mass models}
 
\begin{figure} [t]
	\begin{center}
		\includegraphics[width=8.5 cm]{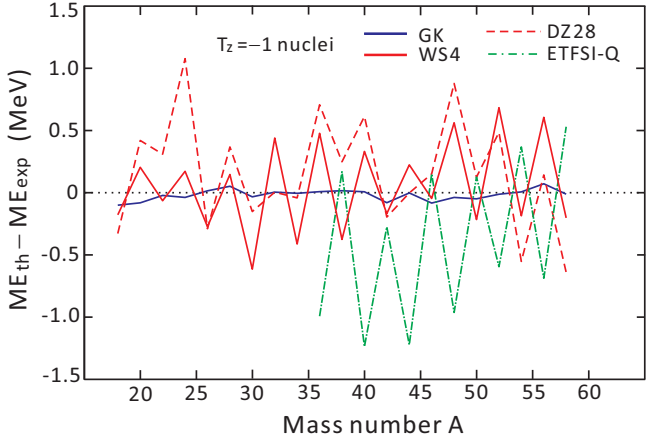}
		\caption{(Colour online) Mass differences between the experimental values and those from different model predictions for the $T_z=-1$ nuclei. 
			\label{Fig11}}
	\end{center}
\end{figure}

The accuracy of the current theoretical nuclear mass models have been recently investigated in Ref.~\cite{Sobi14,Sobi18}. Among the ten often-used models of various nature, the macroscopic-microscopic model of Wang and Liu~\cite{Liu11,Wang11}, the latest version labeled as WS4~\cite{Wang14}, and the Duflo and Zuker (DZ28) mass model~\cite{DZ28} are found to be the most accurate in various mass regions with the smallest RMS (Root-Mean-Squares) values of $250\sim 500$~keV. Their mass predictions are therefore compared with the experimental masses for the $T_z=-1$ nuclei in Fig.~\ref{Fig11}. We also plot the calculated results from the local mass relationships of Garvey and Kelson (GK)~\cite{Gar66} and from the ETFSI-Q mass table~\cite{Pear96}. One can see the predictive powers and accuracies of the models. We noticed that the differences between model predictions and the experimental masses, $ME_{\rm th}-ME_{\rm exp}$, show oscillations for all models. Only the WS4 model yields a regular zig-zag staggering around zero (see Fig.~\ref{Fig11}). This may indicate that both, the nuclear pairing and the smooth $A$-dependence of nuclear masses, are better taken into account
in WS4 than in the other models, leading to more accurate description of the nuclear masses. Of course, the refined treatments of nuclear pairing are still needed in order to reduce the staggering.   
 
The masses of lighter neutron-deficient nuclei can be more precisely predicted by using the local mass relationships such as the GK~\cite{Gar66} mass relation and the IMME. The GK mass relation has been used here to predict the masses of $T_z=-1$ nuclei and compared with the experimental ones in Fig.~\ref{Fig11}. One sees that the simple GK mass relation predicts best the experimental masses compared to the other models, and the regular staggering in the model calculations has been nearly removed in the GK mass predictions.           
 
\subsection{Test of IMME for the $A=52$, $T=2$ multiplet }

In the previous $\beta$-decay studies of $^{52}$Ni~\cite{Faux94,Dossat,Orrigo16}, both $\beta$-delayed protons and $\gamma^{\prime}$s have been observed and a partial decay scheme of $^{52}$Ni was proposed. However, as the masses of $^{52}$Co and its $(2^+)$ isomer had not been measured at that time, the assignments of both the 141-2407 keV $\gamma$ cascade and the 1352-keV protons as from the IAS in $^{52}$Co could not be verified by energy matches. With the precision mass measurements in this work, i.e., the masses of $^{52g,52m}$Co and $^{51}$Fe, we have pointed out~\cite{Xu16} that the former is from the IAS in $^{52}$Co at $E_x=2935(13)$ keV while the latter should be from a $1^+$ state with $E_x=2800(16)$ keV, which could be the analog $1^+$ state identified in its mirror nucleus $^{52}$Mn~\cite{52Mn73,fujita15}. 

\begin{figure} [b]
	\centering
	\rotatebox[]{0}{\includegraphics[width=8 cm,angle=0]{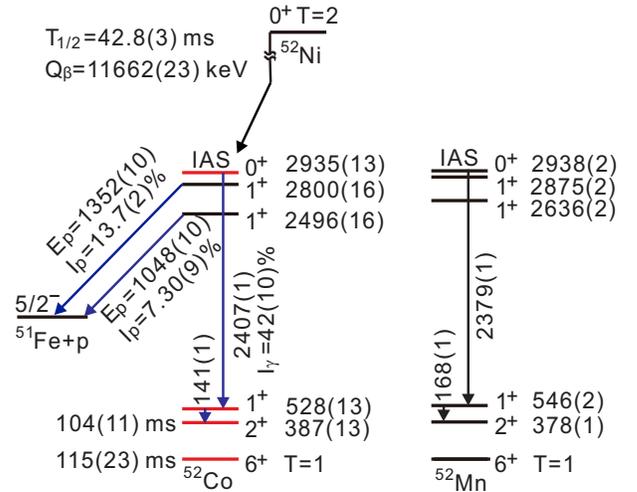}} 
	\caption{
		(Color online) Reconstructed partial decay scheme of $^{52}$Ni compared to that of $^{52}$Mn, where energies are in keV. For $^{52}$Co, the red levels are deduced from the $ME$ values of $^{52g}$Co, $^{52m}$Co (this work), and the $\gamma$-ray energies reported in Ref.~\cite{Orrigo16}; the black ones are determined from the ground state $ME$ of $^{51}$Fe~\cite{AME2012} and the $\beta$-$p$ decay of $^{52}$Ni in Ref.~\cite{Orrigo16}. For $^{52}$Mn, the $1^+$ states and IAS are inferred from Ref.~\cite{fujita15}, while the lowest two levels ($6^+$ and $2^+$ states) and the $\gamma$-transition energies are from Ref.~\cite{52Mn73}, the latter lead to $E_x(\rm IAS)=2925(5)$ keV~\cite{52Mn73}. This inconsistency needs a precise determination for the decay $\gamma$-ray energies in $^{52}$Mn.}
	\label{Fig12}
\end{figure}
We have reconstructed the partial decay scheme of $^{52}$Ni as shown in Fig.~\ref{Fig12}. A partial level scheme of $^{52}$Mn is also given for comparison where the excitation energies are taken from the high-resolution measurement in the $^{52}$Cr($^3$He,t)$^{52}$Mn charge exchange reaction~\cite{fujita15}. The spin-and-parity assignment for the levels of $^{52}$Co is inferred from the analog states of its mirror nucleus $^{52}$Mn~\cite{fujita15}. The main modification in the present level scheme is that we attribute the 1352-keV proton to originate from the decay of the $1^+$ state rather than from the IAS. The excitation energies of the $1^+$ states are calculated by subtracting the g.s. $ME$ value of $^{52}$Co measured in this work from the $ME$ values deduced from the $\beta$-$p$ data. Fig.~\ref{Fig12} shows a good mirror symmetry in the level structure between $^{52}$Co and $^{52}$Mn, and we may consider that the $1^+$ state at $E_x=2800$~keV is most probably the $1^+$ analog state at $E_x=2875$~keV in $^{52}$Mn.

Dossat {\it et al.} have tested the quadratic form of the IMME by adding a cubic term, $d\times T_z^3$, to Eq.~(\ref{eq14}) using $ME({\rm IAS},^{52}{\rm Co})=-31584(18)$ keV deduced from the ground state mass of $^{51}$Fe and the 1352-keV protons. They found that the $d$-coefficient significantly deviated from zero and then attributed this deviation to a misidentification for one of the states assigned to this mass multiplet. Recently, the experimental IAS$^{\prime}$s from $T=1/2$ to $T=3$ have been evaluated and the corresponding IMME coefficients have been investigated~\cite{Mac14}: the IAS assigned in $^{52}$Co~\cite{Dossat,Orrigo16} was excluded in their IMME fit because the $c$-coefficient falls to $\sim 100$ keV from the normal-trend value of 175~keV when the cubic form of IMME was used to fit the data.

\begin{table}
	\caption{Compilation of $ME$ values for ground states (g.s.), Isobaric
		Analog States (IAS$^{\prime}$s) and the corresponding excitation energies
		($E_x$) for $A=52$, $T=2$ quintet. All data are from Ref.~\cite{AME2012} except for $^{52}$Co (this work).
		Also listed are the resultant parameters for quadratic and cubic fits
		(see text). }
	\begin{tabular}{lcccccc} \hline\hline
		Atom &  $T_z$  & $ME$(g.s) & $E_x$ (keV) & $ME$(IAS) \\
		&         & (keV)       &   (keV)          &  (keV)  \\[1mm]
		\hline
		$^{52}$Co & $-1$ & $-34361(8) $ & $ 2935(13) $ & $-31426(10) $  \\
		$^{52}$Fe & $0 $ & $-48332(7) $ & $8557(9)   $ & $-39775(6) $ \\
		$^{52}$Mn & $+1$ & $-50706.9(1.9)$  & $2923(5)$ & $-47784(5)$ \\
		$^{52}$Cr & $+2$ & $-55418.1(6)$    & $ 0 $  & $ -55418.1(6)$ \\
		\hline
		& Quadratic fit: & $\chi_n=1.37 $    &  &  \\
		&  & $a$ (keV)    & $b$ (keV) & $ c $ (keV) \\
		&  & $-39780.8(4.3)$  & $-8179.7(5.6)$ & $ 180.5(2.9) $  \\   
		\hline
		&   Cubic fit:&  $d=5.8(4.2) $ (keV) &  & \\
		&  & $a$ (keV)    & $b$ (keV) & $ c $ (keV) \\
		&  & $-39775(6)$  & $-8184.8(6.7)$ & $ 170(8) $  \\  
		\hline\hline
	\end{tabular}
	\label{table03}
\end{table}
\begin{figure} [t]
	\begin{center}
		\includegraphics[width=7 cm]{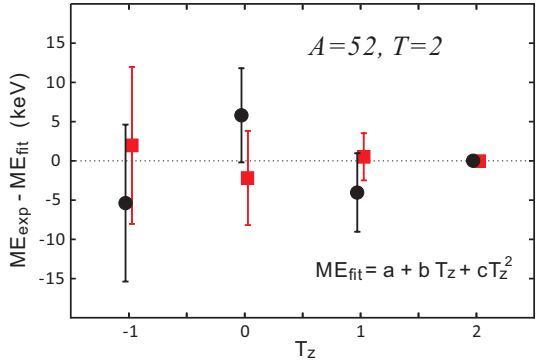}
		\caption{(Colour online) Differences between experimental mass excesses and IMME fit for the $A=52$, $T=2$ multiplet in keV. The mass values in~\cite{AME2012} are used in the fit (black filled circles); the red filled squares correspond to the fit when $E_x({\rm IAS,}^{52}$Mn)=2923(5)~keV~\cite{AME2012} is replaced by $E_x({\rm IAS,}^{52}$Mn)=2938(2)~keV~\cite{fujita15}.        
			\label{Fig13}}
	\end{center}
\end{figure}

Our newly assigned IAS with $ME({\rm IAS},^{52}{\rm Co})=-31426(10)$ keV can be used to test the IMME. For doing this, we compiled the corresponding $ME$ values~\cite{AME2012} of the $A=52, T=2$ multiplet in Table~\ref{table03}. The mass data of the four members for the $A=52, T=2$ multiplet are fitted using the quadratic form of IMME, and the normalized chi-square is obtained to be $\chi_n= 1.37$. 
Fig.~\ref{Fig13} shows the residuals of the fit. We also used the cubic form of IMME to describe the mass data. As given in Table~\ref{table03}, the algebraically calculated $d$-coefficient, $d=5.8(4.2)$~keV, does not significantly deviate from zero when taking the error bars into account. However if $ME$(IAS,$^{52}$Co)$=-31584(18)$~keV is used, the $d$-coefficient is 28.3(4.6)~keV which deviates by $6\sigma$ from zero. We note that the excitation energy of the IAS in $^{52}$Mn has recently been measured to be $E_x({\rm IAS,}^{52}$Mn)=2938(2)~keV in a high-resolution $^{52}$Cr($^3$He,t)$^{52}$Mn charge exchange reaction~\cite{fujita15}. If this value is used in the fitting procedure, a better agreement using the quadratic form of IMME is achieved with $\chi_n=0.45$ (see Fig.~\ref{Fig13}), and $d=-1.7(3.8)$ keV is obtained using the cubic form of IMME.   

\subsection{$p$-$n$ interactions around doubly magic nucleus $^{56}$Ni}

The binding energy of a nucleus $B(Z,N)$ deduced directly from its mass reflects the summed effects of overall interactions inside the nucleus. The various forms of binding-energy differences can be constructed to isolate specific nucleonic interactions. One of such filters is the average interaction strength among the last proton(s) with the last neutron(s) denoted as $\delta V_{pn}$. This $p$-$n$ interaction has been considered to be attractive and closely related to many nuclear structure phenomena such as the onset of collectivity and deformation~\cite{Tal62,Caki09}, changes of underlying shell structure~\cite{Hey85}, and phase transitions in nuclei~\cite{Hey85,Fed77,Fed78}. It has been empirically predicted that the $p$-$n$ interaction strength is sensitive to the spatial overlap of wave functions of the last valence neutron(s) and proton(s)~\cite{Bre06}. Since a big change of the shell model orbitals occurs across some doubly magic nuclei, one expects a large difference of $p$-$n$ interactions crossing the respective shell
closures. Such an idea has been tested in the four quadrants of the nuclear chart surrounding the doubly magic nucleus $^{208}$Pb (see Ref.~\cite{Chen09} and references therein).
The authors have found that the $p$-$n$ interactions are larger when the valence protons and neutrons are both below or above their respective shell closures at $Z=82$ and $N=126$, while the interactions are smaller when one is above and the other is below. 

A similar test of this scenario is now available surrounding the lighter doubly magic nucleus $^{56}$Ni thanks to the experimental masses of $^{56}$Cu and $^{52}$Co measured in this work. 
Here the shell model orbitals jump from $1f_{7/2}$ to $2p_{3/2}$ across the $Z=N=28$ shell closure. In the cases of odd-odd nuclei, the $p$-$n$ interaction strength between the last valence proton and neutron can be calculated in terms of Eq.~(7) of Ref.~\cite{Isa95} 
\begin{equation} \label{eq18}
\begin{split}
&\delta V_{pn}(Z,N)= B(Z,N)+B(Z-1,N-1)\\
&~~~~~~~~~~~~~~~~~~~~-B(Z,N-1)-B(Z-1,N).\\
\end{split}
\end{equation}

\begin{figure} [t]
	\begin{center}
		\includegraphics[width=8cm]{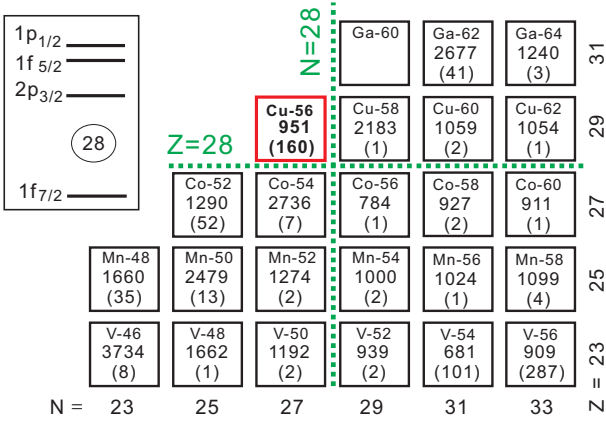}
		\caption{(Colour online) Experimental $\delta V_{pn}$ values in keV for odd-odd nuclei in a small part on the chart of nuclides. The numbers in parentheses are the errors of $\delta V_{pn}$. Four quadrants are defined by the $Z=28$ proton and $N=28$ neutron shell closures. Schematic shell model single-particle orbitals are shown in the left part of this figure. The $\delta V_{pn}$ value for $^{56}$Cu (indicated with red square) is the first experimental result in the upper-left quadrant. It has a similar magnitude as the $\delta V_{pn}$ values in the lower-right quadrant.      	
			\label{Fig14}}
	\end{center}
\end{figure}

The calculated $\delta V_{pn}$ values for some odd-odd nuclei around $^{56}$Ni are given in Fig.~\ref{Fig14}. Note that the $\delta V_{pn}$ value for $^{56}$Cu (indicated with red square) is the first experimental result in the upper-left quadrant, where the large error of its $\delta V_{pn}$ value is due to the uncertainty of the mass of $^{55}$Cu~\cite{AME2012}. One sees from this figure that the $\delta V_{pn}$ values in the lower-left and upper-right quadrants are generally larger than those in the upper-left and lower-right quadrants. This is similar to the lead region and consistent with the expectations that the $p$-$n$ interactions are larger when the valence proton and neutron occupy the same shell model orbital $1f_{7/2}$/($2p_{3/2}$) below/(above) the shell closure at $Z=N=28$, while they are smaller if the proton is in $1f_{7/2}$ and neutron in $2p_{3/2}$ or vice versa. The $\delta V_{pn}$ values of $N=Z$ nuclei are the largest due to the  Wigner's SU(4) symmetry~\cite{Isa95} that enhances the overlap of the wave functions. 

Establishing a general feature of $\delta V_{pn}$ throughout the nuclear chart has important impact on the mass predictions using local mass relationships~\cite{Gar66,Jane72,Bao13,Cheng14}. It is believed that the Garvey-Kelson mass relations~\cite{Gar66} or the recently reconstituted ones~ \cite{Bao13,Cheng14} are able to predict the unknown masses with higher accuracy compared to the mass-model calculations (see Fig.~\ref{Fig11} for example). It is worthwhile to note that these local mass relations can be derived under the condition that the $\delta V_{pn}$ values deduced from Eq.~(\ref{eq18}) are identical for two $neighboring$ nuclei. Here by $neighboring$ it means the two $neighboring$ nuclei of same parity (even-even, even-odd, odd-even, and odd-odd) along an isotopic, isotonic, isobaric, and isospin projection chain. The local mass relations in Refs.~\cite{Bao13,Cheng14} are equivalent to GK formulas when no same-parity of $neighboring$ nuclei is required (for details, the reader is referred to Fig.~2 in Ref.~\cite{Cheng14}). 

\begin{figure} [t]
	\begin{center}
		\includegraphics[width=6 cm]{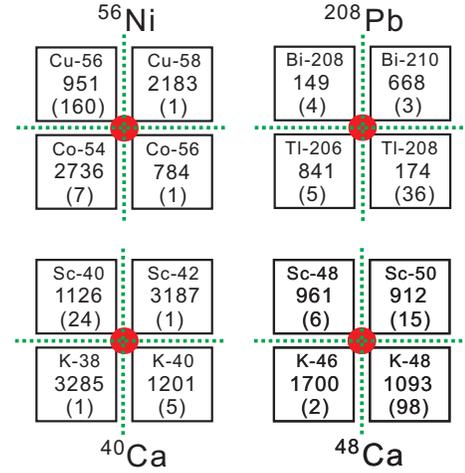}
		\caption{(Colour online) Same as Fig.~\ref{Fig14} but for four odd-odd nuclei around the doubly magic nuclei $^{40}$Ca, $^{48}$Ca, $^{56}$Ni, and $^{208}$Pb marked with red filled circles.       	
			\label{Fig15}}
	\end{center}
\end{figure}
  
Fig.~\ref{Fig15} shows available $\delta V_{pn}$ values calculated according to Eq.~(\ref{eq18}) for four odd-odd nuclei around $^{40}$Ca, $^{48}$Ca, $^{56}$Ni, and $^{208}$Pb. One sees that the $\delta V_{pn}$ values change suddenly across the shell closures which could be due to sudden changes of the spatial overlap of shell model orbitals of the last valence neutron and proton. Such a sudden change of $\delta V_{pn}$ values are observed along the isotopic and isotonic chains as well as along the isospin projection chain around $^{48}$Ca. This behavior may set constraints on the reliability of the local mass relations. For example, in lead region, the mass of $^{207}$Hg is predicted to be $\sim 500$ keV less bound than the experimental value using Eqs.~(8) and (9) of Ref.~\cite{Cheng14}. And the masses of $^{50,52}$Sc are predicted to be $800\sim 900$ keV more bound than the experimental ones using Eq.~(10) of Ref.~\cite{Cheng14}. 

A close inspection of Figs.~\ref{Fig14} and~\ref{Fig15} reveals that the prerequisite of identical $\delta V_{pn}$ values holds well for isobars surrounding the four doubly magic nuclei, and a mirror symmetry of $\delta V_{pn}$ along the $Z=N$ line is observed in the lower-left quadrant of Fig.~\ref{Fig14}. This indicates that the transverse GK mass relation and Eq.~(10) of Ref.~\cite{Cheng14} can be used in these cases. Indeed, the mass of $^{207}$Hg is calculated using the Eq.~(10) of Ref.~\cite{Cheng14} to be nearly 25~keV lower than the experimental value. In the same way, the mass excess of $^{55}$Cu could be predicted to be $-31802(16)$~keV using the precise masses of $^{56}$Cu and $^{54}$Ni measured in this work. This value agrees well with the IMME prediction of $ME(^{55}{\rm Cu})=-31782(4)$~keV. Both predicted $ME$ values for $^{55}$Cu are 
$\sim 150$~keV lower than the experimental value of -31635(156) keV~\cite{Yan13}. In this sense, a more precise mass measurement for $^{55}$Cu would be desired.  


\section{Summary and Conclusions }
We have performed isochronous mass measurements for the neutron-deficient $fp$-shell nuclei produced via projectile fragmentation of an energetic $^{58}$Ni beam. Special techniques have been applied to the current measurements and data analyses in order to increase the resolving power of isochronous mass spectrometry in the heavy ion storage ring CSRe in Lanzhou, e.g., inserting a metal slit in the dispersion section of the ring, and using a new technique to correct the effects of the unstable magnetic fields of the RIBLL2-CSRe system. On the basis of the newly measured masses, several nuclear structure studies in the $fp$ shell have been performed. Main results and conclusions from this work are summarized as follows:
\par
(1) The mass excesses of the $T_z=-1$ nuclei $^{52g,52m}$Co and $^{56}$Cu have been measured for the first time in this experiment with an uncertainty of $\sim 10$ keV. This is the highest precision reached in the isochronous mass spectrometry for short-lived neutron-deficient nuclei. Our measurements show that $^{52g,52m}$Co and $^{56}$Cu are $\sim 370$~keV and $\sim 400$~keV more bound than the evaluations in AME$^{\prime}$12, respectively. The new mass of $^{56}$Cu allows us to observe the mirror symmetry of low-spin excited levels between $^{56}$Cu and $^{56}$Co within an uncertainty of 50~keV. The energy of the $T=2$ IAS in $^{52}$Co is newly assigned precisely, which fits well into the fundamental Isobaric Multiplet Mass Equation.    
\par
(2) The mass excesses of five $T_z=-1$ nuclei $^{44}$V, $^{46}$Cr, $^{48}$Mn, $^{50}$Fe, and $^{54}$Ni have been re-measured, the precision of which except for $^{46}$Cr is one order of magnitude higher than the values in AME$^{\prime}$12. Especially, the mass excesses of $^{44}$V and $^{54}$Ni are $\sim 300$~keV and $-60$~keV, respectively, deviating from the literature ones. The new mass data allow us to establish the general $A$ dependent features of vector and tensor Coulomb energies up to $A=58$ for the $T=1$ isobaric triplets. We have shown that the oscillation pattern of tensor Coulomb energy persists for $fp$-shell nuclei. This fact may provide a test ground for investigating the effects of isospin symmetry breaking, as well as a guideline for mass extrapolation and measurement of heavy nuclei in and even beyond $fp$ shell.  
\par
(3) The masses of four $T_z=-1/2$ nuclei $^{45}$V, $^{47}$Cr, $^{49}$Mn, and $^{51}$Fe, which are obviously outside the isochronous window, were also measured. The deduced mass excess values agree well, within the experimental errors, with the recent JYFLTRAP measurements (for $^{45}$V and $^{49}$Mn) or with our previous IMS measurements (for $^{47}$Cr and $^{51}$Fe) in CSRe. The consistent results for $^{51}$Fe and the new mass of $^{52m}$Co help us to re-assign the $T=2$ IAS in $^{52}$Co by referring to the experimental data on $\beta$-delayed protons and $\beta$-delayed $\gamma^{\prime} $s of $^{52}$Ni.
\par
(4) The mass excess of the expected low-lying $(6^+)$ isomer in $^{44}$V has been determined for the first time in this experiment to be $-23541(19)$~keV, which is, similar to its ground state, $\sim 300$~keV less bound than the evaluations in AME$^{\prime}$12. The excitation energy $E_x=286(28)$~keV was found to be very close to the $E_x$ value of an analog state ($E_x=271$~keV) in its mirror nucleus $^{44}$Sc.
\par            
(5) We have investigated the $Z$ and $N$ dependences of residual $p$-$n$ interactions around the doubly magic nucleus $^{56}$Ni using our new mass of $^{56}$Cu. Similar to the case around $^{208}$Pb, the hypothesis still holds that the $p$-$n$ interaction strength is positively correlated with the spatial overlap of wave functions of the last valence neutron(s) and proton(s). Further analyses show that the empirical $p$-$n$ interactions deduced from atomic masses change
suddenly across the shell closures throughout the chart of nuclides. 
This is due to sudden changes of the spatial overlap of shell model orbitals of
the last valence neutron and proton. Consequently this sets constraints on the applicability of local mass relationships, e.g., Eqs.~(8), (9) and (10) of Ref.~\cite{Cheng14}, to predict unknown masses with high accuracy.   

\acknowledgments 
We thank the staffs in the accelerator division of IMP for providing stable beam. This work is supported in part by the Major State Basic Research Development Program of China (Contract No. 2013CB834401), the National Key Program for S$\&$T Research and Development (Contract No. 2016YFA0400504), the Key Research Program of Frontier Sciences, CAS (Grant No. QYZDJ-SSW-SLH005), and the NSFC grants U1232208, U1432125, 11605248, 11605249, 11605252, 11605267, 11575112, 11575007, the European Research Council (ERC) under the EU Horizon 2020 research and innovation programme (ERCCG 682841 "ASTRUm").  
Y.H.Z. acknowledges support by the ExtreMe Matter Institute EMMI at the GSI Helmholtzzentrum f{\"u}r Schwerionenforschung, Darmstadt, Germany.
Y.A.L. is supported by CAS visiting professorship for senior international scientists (Grant No. 2009J2-23), the CAS External Cooperation Program (Grant No. GJHZ1305), HGF-CAS joint research group (HCJRG-108). 
K.B. acknowledge support by the
Nuclear Astrophysics Virtual Institute (NAVI) of the Helmholtz Association. 


\end{document}